\documentstyle[12pt]{article}

\begin{document}

\vskip .5in
\title{\bf
Integrable Perturbations of $W_n$ and WZW Models
}
\author{
{\bf
I.Vaysburd
}\\
\normalsize SISSA, via Beirut, 2-4, I-34013 Trieste,
{\bf Italy}}
\newpage
\maketitle
\begin{abstract}
We present a new class of 2d integrable models obtained as perturbations
of minimal CFT with W-symmetry by fundamental weight primaries.
These models are generalisations of well known $(1,2)$-perturbed
Virasoro minimal models.
In the large $p$ (number of minimal model) limit they coincide with scalar
perturbations of WZW theories. The algebra of conserved charges is discussed
in this limit. We prove that it is noncommutative and coincides
with twisted affine algebra $G$ represented in a space of asymptotic
states.
We conjecture that scattering in these models for generic $p$
is described by $S$-matrix of the $q$-deformed
$G$ - algebra with $q$ being root of unity.

\end{abstract}
\newpage
\section{Introduction}

It was pointed out in \cite{zam} that a large class of  2d quantum
integrable field theories can be considered as certain perturbation
of 2d CFT by an operator $\Phi (z,\bar{z})$ from
the Hilbert space of a theory
\begin{equation}
{S_{IM}=S_{CFT}+\lambda\int\Phi(z,\bar{z})d^2z}
\label{eq:1}
\end{equation}
If the conformal symmetry generated by holomorphic
tensor $T(z) (\bar{T}(\bar{z}))$ is the only affine symmetry of a
theory, the field $\Phi$ has to be completely degenerate primary
with respect to this symmetry (see \cite{Mussardo} for review).
Complete degeneracy leads to a substantial lowering of dimensionality
of irreducible highest weight representation. As a result,
some holomorphic polynomials of $T(z)$  and it's derivatives survive
the perturbation. They become left components of infinitely many integrals
of motion (IM).
Strictly speaking, aside from dimensionality of irreducible representation
of Virasoro algebra defined by perturbing field, one has to take into
account renormalizability properties of a perturbed CFT. Renormalization
generically brings about counterterms which are fields
from a fusion algebra generated by $\Phi$ and their conformal descendants.
These counterterms could destroy integrability. They do not appear if
$\Phi$ is the most relevant primary in a theory, or at least in the fusion
algebra generated by $\Phi$. But even if this is not the case
one can argue \cite{zamol} that counterterms (descendants of other
primaries) do not contribute to renormalization of conservation laws
unless some special "resonance" condition on conformal dimensions of
primaries holds. We will keep in mind the case when "resonance"
condition does not spoil integrability.
An examination of the Virasoro characters of irreducible highest weight
representations shows all of the possible relevant primary
fields leading to integrable theories.
These are $\Phi_{(1,2)}[\Phi_{(2,1)}], \Phi_{(1,3)}, \Phi_{(1,5)}$.
The field $\Phi_{(1,5)}$ turns out to be relevant
in nonunitary mimimal models $(p,q)$ if $p<q/2$, $q>6$.
 The field $\Phi_{(2,1)}$ is relevant for $q/2<p<q$.
This (2,1) and (1,5) duality is probably the simplest example of
Affine Toda dualities which has been observed recently in several papers
\cite{Gri,corrigan,watts}.
In this particular case we deal with $A_2^{(2)}$ selfduality.
This does not mean that these are the only integrable perturbations
for all the $(p,q)$ - models. For some
of them perturbations not belonging
to this list are integrable, but operators from the list have a special
reason to be integrable, namely, they can be
assosiated with quantum Affine Toda Theories (ATT) \cite {EY,HM,FF1,FF2,FF3}.
The field $\Phi_{(1,3)}$ corresponds to the complex $A_1^{(1)}$, or
the Sine-Gordon model,
$\Phi_{(1,2)},\Phi_{(2,1)},\Phi_{(1,5)}$ lead to the complex $A_2^{(2)}$ ATT
\cite{IzKor}. The latter one is non-hermitian and admits a physical
interpretation after an appropriate quantum group (QG) reduction \cite{
Smirnov}.

The same argument applied to $W_3$ theories gives the following list of
operators $
F_{(22\vert 11)}, F_{(21\vert 11)},
F_{(11\vert 21)}, F_{(41\vert 11)}
$, where we use notations introduced by Zamolodchikov, Fateev and Lukyanov
\cite{ZFW,FLW}.
The first operator corresponds to $A_{2}^{(1)}$. It is well known
$(1,1,Adj)$- perturbation explicitly studied in \cite{FdV}. It preserves
$S_3$ symmetry of $W_3$ minimal model and is in a sense an analog of
$(1,3)$ from Virasoro minimal model.
The second and the third operators correspond to $G_2^{(1)}$ ATT and
are generalizations of (1,2) and (2,1). The last operator corresponds
to $D_4^{(3)}$ which is dual to $G_2^{(1)}$ and generalizes (1,5).
$A_2$ finite Dynkin diagram enters as a subdiagram affine diagrams
$A_2^{(1)}$, $G_2^{(1)}$ and $D_4^{(3)}$.

For $W_4$ algebra we have
$(212\vert 111),
(121\vert 111),
(131\vert 111)$ corresponding to
$A_3^{(1)}, B_3^{(1)}$ and $A_5^{(2)}$.

These simple observations are just hints and possible integrability of
listed perturbations needs to be proven rigorously.
In Sections 2,3 we prove integrability of $(21\vert 11)$, $(11\vert 21)$
perturbation of minimal $W_3$ $(p,p+1)$ models and
$(121\vert 111)$ , $(111\vert 121)$ perturbations of $ W_4$.
In Section 4 we argue that in the $p\to\infty$ limit factorized
scattering theory defined by these perturbations is one of $A_2^{(1)}$
 $k=1$ WZW
model perturbed by $(1,0)\otimes\overline{(0,1)}$ operator and one of
$A_3^{(1)}$ $k=1$ WZW model perturbed by $(0,1,0)\otimes\overline{(0,1,0)}$
operator. These perturbations preserve diagonal $A_2$ and $A_3$
symmetries respectively.
It appears to be possible to construct infinite set of
conserved non-commutative charges in these perturbed WZW acting as
$D_4^{(3)}$ and $A_5^{(2)}$ on asymptotic states. It is interesting
what is the full algebra generated by these charges. The work on it is
in progress now.
It seems natural to assume that for finite $p$ these theories
exhibit $q$-deformed $D_4^{(3)}$ and $A_5^{(2)}$ symmetries, i.e.
fundamental kinks belong to vector representations of twisted
affine Lie algebras.

\section {Integrable perturbations of $W_3$ minimal models.}

$(21\vert 11)$ and $(12\vert 11)$ operators form  $S_3$-doublet
and can be represented by free scalar fields:
\begin {equation}
F_{(21\vert 11)} = e^{-i\alpha_+\vec{\omega_1}\vec{\phi}} = \Omega(z,\bar z)
\label {eq:2}
\end{equation}
\begin{equation}
F_{(12\vert 11)} = e^{-i\alpha_+\vec{\omega_2}\vec\phi} = \Omega^+(z,\bar z)
\label {eq:3}
\end{equation}
$\omega_1$, $\omega_2$ are fundamental weights of su(3) algebra
and $\alpha_+ = \sqrt{p\over {p+1}}$.
In $(p,p+1)$ model their dimension equals
\begin {equation}
\Delta^{(p)}_{(21\vert 11)} = \Delta^{(p)}_{(12\vert 11)} =
{1\over 3} (1-{4\over{p+1}})
\label {eq:4}
\end {equation}
IM of  CFT, i.e. normally ordered polynomials of Lorentz spin $s$
 $P_s(T,W)$ are generically spoiled by $W$-descendants of $\Omega(z,\bar z)$
but it may happen that for some of them this $W$-descendant is total
derivative:
\begin {equation}
\bar{\partial} P_s(T,W) = \partial \Omega_{s-2}(z,\bar z)
\label {eq:5}
\end {equation}
That is always the case
when a number of level $s$ descendants excluding total derivatives
in a vacuum module of a theory is greater than a number
of level $s-1$ descendants of perturbing field.
Comparison of $W_3$ characters $\chi^{(p)}_{(21\vert 11)}$ and
$\chi^{(p)}_{(11\vert 11)}$ with total derivatives subtracted
shows that it happens for $s=6$ and any $p$.
Formulae for $W_n^{(p)}$ characters obtained in \cite{FatLuk}
reads:
\begin{equation}
\chi^{(p)}(\Omega_1,\Omega_2) = [\prod\limits_{i=1}^\infty (1-q^i)]^{-r}
\sum_{s\in w}\sum_{\lambda\in\Lambda_R}det(s)q^{[ps(\Omega_1)-
(p+1)\Omega_2 + p(p+1)\lambda]^2/2p(p+1)}
\label{eq:6}
\end{equation}
where the sums run over the elements $s$ of the Weyl group $w$
and the root lattice $\Lambda_R$ of the Lie algebra $A_{n-1}$.
Let us note that Verma module of completely degenerate highest weight of
$W_n$ algebra contains $n$ independent null-vectors. The sum over
Weyl group accounts for $(n-1)$ of them as long as the sum over root lattice
accounts for $n$-th one. If $p$ is sufficiently large $n$-th null-vector
does not affect multiplicities of first levels and is irrelevant for
our considerations. So we can keep only the $\lambda = 0$ term. Sum
over the Weyl group can be calculated explicitly. Simplified character
formula  for $W_3$ is
\begin{equation}
\tilde{\chi}^{(p)}_{(21\vert 11)} = q^\Delta
[\prod\limits_{i=1}^\infty (1-q^i)]^{-2}(1-q-q^2+q^4+q^5-q^6)
\label{eq:7}
\end{equation}
where $\Delta$ means Virasoro dimension of primary field.
Subtraction of total derivatives yields
\begin{equation}
(1-q)\tilde{\chi}^{(p)}(\Omega_1,\Omega_2) = q^\Delta\sum_i D_i^{(\Omega_1,
\Omega_2)} q^i
\label {eq:8}
\end{equation}
where $D_i$ are multiplicities we search for. For $p>5$ (7) and (8) give
$$
D_6^{(11\vert 11)}=4
$$
$$
D_5^{(21\vert 11)}=3
$$
For $p=4$ (critical $Z_3$ model) additional null-vectors are to be taken
into account what lowers by one both dimensionalities
$$
D_6^{(11\vert 11)}=3
$$
$$
D_5^{(21\vert 11)}=2
$$
It means that there exists
\begin{equation}
P_6(z,\bar z)= a:T^3: + b:(\partial T)^2: + c:W^2: + d:W\partial T:
\label{eq:9}
\end{equation}
obeying (5). Explicit expressions for $a,b,c,d$ can be obtained but
they are not essential for the rest of the paper. Let us note nevertheless
that $c$ vanishes for the first minimal model $W_3^{(4)}$.
Appearance of the IM of spin 5
\begin{equation}
Q_5=\int dz P_6(z) + \int d\bar z \Omega_4(z, \bar z)
\label{eq:10}
\end{equation}
does not seem strange. The weight of perturbing operator together
with two simple roots of $A_2$ algebra forms the root system of affine
$G_2^{(1)}$ algebra with Coxeter exponents 1,5 (mod 6). So, the
nontrivial IM we have found confirms $G_2^{(1)}$-origin of
$(21\vert 11)$ perturbation of $W_3$.
The same reasoning is valid for $\Omega^+$- perturbation with exchange
$d\to -d$.
Let us note, that $s=5$ conserved charge can be constructed for
$(11\vert 21)$ perturbation as well. But $G_2^{(1)}$-structure
shows up only for sufficiently large $p$. For the lowest $p=4$
$$
(11\vert 21) = \Psi (z)
$$
$$
(11\vert 12) = \Psi^+ (z)
$$
these are Zamolodchikov-Fateev parafermions. The theory perturbed
by hermitian combimation of them possess IM with $s=3$ due to additional
degeneracy \cite{FatPar} and coincides with restricted SG.

\section{Integrable perturbation of $W_4$ minimal model}

Besides operator $(1,1,Adj)$ in $W_4^{(p)}$ models exists one more
defining integrable perturbation
\begin {equation}
F_{(121\vert 111)} = e^{-i\alpha_+\vec{\omega_2}\vec{\phi}}
\label{eq:11}
\end{equation}
$\omega_2$ is the second fundamental weight of $A_3$.
In order to prove it we have to compare two characters
$\chi ^{(p)}_{(121\vert 111)}$ and $\chi ^{(p)}_{(111\vert 111)}$.
As before we will consider several lowest levels of Verma modules
unaffected by fourth null-vector. Then the same approximation
as in the previous case is applicable.
It becomes exact for irrational values of central charge and for
$c=n-1=3$.
\begin{equation}
\tilde{\chi}_{(121\vert 111)} = q^\Delta
[\prod\limits_{i=1}^\infty (1-q^i)]^{-3}(1-2q+2q^4+2q^5-3q^6-3q^7+2q^8+2q^9
-2q^{12}+q^{13})
\label{eq:12}
\end{equation}
where $\Delta$ means Virasoro dimension of primary field.
Subtraction of total derivatives gives
$$
D_4^{(111\vert 111)}=2
$$
$$
D_3^{(121\vert 111)}=1
$$
So we discover a nontrivial IM of spin $s=3$
$$
P_3 = \int dz (:T^2:+ kW_4) + \lambda\int d\bar{z}F_2(z,\bar z)
$$
The weight of perturbing operator along
with three simple roots of $A_3$ algebra forms the root system of affine
$B_3^{(1)}$ algebra with Coxeter exponents 1,3,5 (mod 6). So, the
nontrivial IM we have found confirms $B_3^{(1)}$-origin of
$(121\vert 111)$ perturbation of $W_4$.
We conjecture that these integrable models are lowest in  hierarchy
$WD_n^{(p)}+F_{vect}$, where $F_{vect}$ is $(21\dots 1\vert 1\dots 1)$ -
primary.

\section{$p\to\infty$ and perturbed WZW}

Let us consider $p\to\infty$ limit of perturbed $WX_n$ models.
Minimal $WX_n$ CFT constructed out of simply-laced affine Lie algebra
$$
WX_n^{(x(n)+p)} = {X_{n,1}^{(1)}\times X_{n,p}^{(1)}\over X_{n,p+1}^{(1)}}
$$
admit well known integrable $(1,1,Adj)$ perturbation. It coincides in this
limit
with $J\bar J$- perturbation of $X_{n,1}^{(1)}$ WZW model, or with complex
ATT at some specific value of coupling constant. It
enjoys $X_n$ symmetry which acts as diagonal
$(J_0 + \bar J_0)$- algebra of zero modes. Solitonic particles appear as
fundamental $n$-plet of $X_n$. As was argued in \cite {Leclair,
BernLec,Smir,SmirResh, EguYang} for
$p<\infty$ this symmetry becomes $q$-deformed quantum group with
$$
q^{x(n)+p} = 1,
$$
where $x(n)$ is a dual Coxeter number of $X_n$.
We argue that for fundamental weight perturbation
 we get some affine twisted symmetry with $X_n$ as zero mode
subalgebra. Let $X_n$ be $A_1$. For $(1,2)$ perturbation of Virasoro minimal
models it was shown by F.Smirnov \cite {Smirnov} that for $p\to\infty$
scattering theory is one of SG model at $\beta = {1\over \sqrt 2}$. Let us
note that this model can be thought of as a fundamental weight perturbation
of $A_{1}$ $k=1$ WZW theory presreving diagonal $A_1$.
$$
S^{(p)}_{(1,2)}\vert_{p\to\infty} = S_{wzw} + \lambda\int d^2z
[\phi^{1/2,1/2}\otimes \overline{\phi}^{1/2,-1/2} -
\phi^{1/2,-1/2}\otimes \overline{\phi}^{1/2,1/2}]
$$

\noindent
One might think that being simply-laced SG model "forgets" about it's
$A_2^{(2)}$ - origin, but amazingly it turns out to keep "memory"
of $A_2^{(2)}$. Namely, it exhibits noncommuting conserved charges
which act on solitonic sector by vector representation of $A_2^{(2)}$.
SG theory at generic point possess commuting IM with spins
$s=1,3,5\dots $. At $\beta = 1/\sqrt 2 $ ( the point in consideration)
there are additional IM. In order to classify them let us note that
those are first of all $s=0$ charges, generators of $A_1$.
\begin{equation}
L^{\pm,0} = J^{\pm,0}_0 + \bar J^{\pm,0}_0
\label{eq:13}
\end{equation}
Each IM $Q_s$ of odd (possibly negative) spin aquires an angular
 momentum with respect to this $A_1$ algebra. It turns out that

\vskip .2in \hskip 2in
$j=2$ for $s=6k+3$,

\vskip .2in \hskip 2in
$j=0$ for $s=6k\pm 1$.
\vskip .2in
\noindent
Besides it there are additional $j=1$ charges of Lorentzian spin $s=6k$,
 $k\in Z$.
Non-zero $j$-momentum means that there is a $j$-multiplet of charges
of a spin $s$. Let us prove this result for $s=3$.
We deal with perturbed WZW model, i.e. original space of holomorphic
currents is given by polynomials of $A_1^{(1)}$ currents
$P^{j,m}_s(J^{\pm, 0}(z))$ and
$\bar P^{j,m}_s(\bar J^{\pm, 0}(\bar z))$.
$j,m$ are angular momentum and it's $0$- component of
$P$-charges with respect to $A_1$ symmetry. After perturbation
we get
\begin{equation}
\bar \partial P_s^{j,m}= \lambda
[\phi^{l,m+1/2}_{s-1}\otimes \overline{\phi}^{1/2,-1/2} -
\phi^{l,m-1/2}_{s-1}\otimes \overline{\phi}^{1/2,1/2}]
\label{eq:14}
\end{equation}
$\phi^{j,m}_s$ is a field in Verma module of highest weight $1/2$
with given quantum numbers. $A_1$-momentum conservation yields
$$
j=l\pm 1/2
$$
Now we are going to apply counting argument invented by A.Zamolodchikov
with some modification. Instead of counting of multiplicities of
single operators we will count multiplicities of $A_1$-multiplets.
Examining characters of affine $A_1^{(1)}$ algebra with $k=1$
one notices that after subtraction of total derivatives
in the vacuum module at $s=4$ two multiplets $(j=0,2)$  remain there.
In $1/2$-module at $s=3$ there is only one $l=1/2$ which fails to
obey $j=l\pm 1/2$ selection rule for $j=2$. So, we stay along with
five conserved charges of spin $s=3$!
\begin{equation}
Q^{2,m}_3 = \int dz P^{2,m}_4 + \lambda\int d\bar z
[\phi^{3/2,m+1/2}_2\otimes \overline{\phi}^{1/2,-1/2} +
\phi^{3/2,m-1/2}_2\otimes \overline{\phi}^{1/2,1/2}]
\label{eq:15}
\end{equation}
In this formula we imply that $\phi^{l,m}=0$ if $m>l$.
$P^{2,m}_4$ is the only $j=2$-plet in $0$-module on level 4
and $\phi^{3/2,m}_2$ is the only $j=3/2$-plet in $1/2$-module
on level 2. So, (15) unambiguously defines conserved charges.
These five charges along with their right counterparts of
$s=-3$ and three $A_1$-charges generate the whole algebra.
Consider now the $A_2^{(2)} \mapsto A_2^{(1)}$ embedding.
$A_2$ can be thought of as $Z_2$-graded algebra
$$
A_2=A_2^+ + A_2^-,
$$
where $A_2^+ = A_1 = so(3)$ and $A_2^-$ is $j=2$ representation
of $so(3)$ adjoint action.
Commutation relation in $A_2$ can be schematically written down
as follows

$$
[A_2^+,A_2^+]=A_2^+
,\hskip .5in
[A_2^+,A_2^-]=A_2^-
,\hskip .5in
[A_2^-,A_2^-]=A_2^+.
$$

\noindent
Now we take the  subalgebra of $A_2^{(1)}$
generated by
$
[(A_2^+)_{2n}$, $(A_2^-)_{2n+1}$; $n\in Z]$

\noindent
This subalgebra coincides with $A_2^{(2)}$ for $k=0$.
Conserved $Q$-charges can be identified with generators of $A_2^{(2)}$

$$
Q^{2,m}_{6n+3}=(A_2^-)_{2n+1},\hskip .3in
Q^{1,m}_{6n}=(A_2^+)_{2n},\hskip .3in
n\in Z,\hskip .3in n\geq 0.
$$

\noindent
If action of $Q$-charges on asymptotic triplet states (kink-first breather-
antikink) $\vert \theta ,\pm >, \vert \theta ,0>$
is considered, one can complete
the identification by

$$
\bar Q^{2,m}_{6n+3}=(A_2^-)_{2n+1},\hskip .3in
\bar Q^{1,m}_{6n}=(A_2^+)_{2n},\hskip .3in
 n\in Z,\hskip .3in n<0.
$$

\noindent
This identification immediately follows if we recall that asymptotic states
are eigenstates of canonical IM of SG model -  $Q^{2,0}_s,
\bar Q^{2,0}_s$ in our notations -  and
form $A_1$-triplet. So the explicit action of conserved charges
on asymtotic states after an appropriate normalization
is given by $3\times 3$ matrices of $A_2^{(2)}$ vector
representation
$
[A_2^+ \tau^{2n}$, $A_2^- \tau^{2n+1}$, $n\in Z]
$,
$\tau = e^{3\theta}$, $\theta$ - particle rapidity.

\noindent
A very similar picture can be obtained for $(21\vert 11)$-perturbed
$W_3^{(p)}$ minimal models. In the $p\to\infty$ limit it coincides
 with $A_2^{(1)}$ $k=1$ WZW model perturbed by $A_2$-scalar operator
$$
H=\Phi^{(1,0)}\otimes\bar \Phi^{(-1,0)}+\Phi^{(-1,1)}\otimes\bar \Phi^{(1,-1)}+
\Phi^{(-1,-1)}\otimes\bar \Phi^{(1,1)}
$$
Upper indices indicate $A_2$-weights of operators in fundamental and
conjugate representations.
This model is nothing but complex $A_2$ ATT at $\beta=1/\sqrt 3$ and possess
therefore a set of commuting IM of $s=1,2$ (mod $3$). One can check that these
IM are not singlets but rather multiplets of diagonal $A_2$ symmetry
preserved by $H$-perturbation. We will list the highest weights $(j_1,j_2)$
of these multiplets

\vskip .2in \hskip 1.5in
$(j_1,j_2)=(3,0)$ for $s=6n+2,$

\vskip .2in \hskip 1.5in
$(j_1,j_2)=(0,3)$ for $s=6n-2,$

\vskip .2in \hskip 1.5in
$(j_1,j_2)=(0,0)$ for $s=6n\pm 1,$

\vskip .2in \hskip 1.5in
$n\in Z.$

\vskip .2in
\noindent
Besides these ones there are $s=6n$ charges absent in generic $A_2^{(1)}$
ATT with $(j_1,j_2)=(1,1)$.
This result can be obtained by modified counting argument as in the previous
case. We argue that conserved charges $Q^{(j_1,j_2)}_s$ and
$\bar Q^{(j_1,j_2)}_{-s}$ for $s=2n$ form two subalgebras of $D_4^{(3)}$
and give the full $D_4^{(3)}$ if acting on solitonic sector of the theory.
Let us define first $D_4^{(3)}$ as a certain subalgebra of $D_4^{(1)}$.
$D_4$ can be decomposed into direct sum of three linear spaces:
$$
D_4 = A^{(1,1)}\oplus A^{(3,0)}\oplus A^{(0,3)}.
$$
These are spaces of irreducble representations of adjoint
$A_2$-action embedded in $D_4$ by $(1,1)$-representation.
Commutation relation in $D_4$ schematically look as follows
$$
[A^{(1,1)},A^{(1,1)}]=A^{(1,1)},\hskip .3in
[A^{(1,1)},A^{(3,0)}]=A^{(3,0)},\hskip .3in
[A^{(1,1)},A^{(0,3)}]=A^{(0,3)},
$$
$$
[A^{(3,0)},A^{(3,0)}]=A^{(0,3)},\hskip .3in
[A^{(3,0)},A^{(0,3)}]=A^{(1,1)},\hskip .3in
[A^{(0,3)},A^{(0,3)}]=A^{(3,0)}.
$$
Thereby $Z_3$-graded Lie algebra structure is defined.
Then $D_4^{(3)}$ basis is subset of $D_4^{(1)}$ one's

$$
Q_{6n}^{(1,1)}=A^{(1,1)}_{3n},\hskip .2in
 Q_{6n+2}^{(3,0)}=A^{(3,0)}_{3n+1},\hskip .2in
Q_{6n-2}^{(0,3)}=A^{(0,3)}_{3n-1},\hskip .2in n\geq 0
$$

$$
\bar Q_{6n}^{(1,1)}=A^{(1,1)}_{3n},\hskip .2in
\bar Q_{6n+2}^{(3,0)}=A^{(3,0)}_{3n+1},\hskip .2in
\bar Q_{6n-2}^{(0,3)}=A^{(0,3)}_{3n-1},\hskip .2in n\leq 0.
$$

\noindent
The $S$-matrix of complex $A_n^{(1)}$ ATT has been constructed recently
by T.Hollowood \cite {Hollow}. In $A_2^{(1)}$ ATT at $\beta = 1/\sqrt 3$
scattering bootstrap exibits eight lightest particles of the same mass
including 3 and $\bar 3$-plets of fundamental kinks and antikins along
with two breathers. The mass of these breathers coincides with the mass
of kinks precisely at that value of $\beta$! All these particles form
vector representation of $D_4^{(3)}$.
It is curious that here we encounter kind of non-simply-laced duality
observed recently in \cite {Gri,corrigan}.
Namely, starting from $G_2^{(1)}$-like perturbation of $W_3$ we get in
the $p\to\infty$ limit hidden $D_4^{(3)}$ which can be obtained by
inversion of the $G_2^{(1)}$ Dynkin diagram arrows. The same is true
for SG but $A_2^{(2)}$ hidden symmetry is selfdual.
Now it looks natural to expect that in $W_4^{(p)}$ model perturbed
by $B_3^{(1)}$-like $(121\vert 111)$-operator we should find $A_5^{(2)}$
dual to $B_3^{(1)}$. Let us argue that it is true.
As before in the $p\to\infty$ limit we arrive at perturbed $A_3$ WZW
model with $k=1$:
\begin{equation}
S=S_{wzw}+\lambda\int d^2z V(z,\bar z)
\label{eq:100}
\end{equation}
where
$$
V(z,\bar z)=\sum_{\vec\mu\in\pi_{(0,1,0)}}
\Psi ^{(0,1,0)}_{\vec\mu}\otimes\bar\Psi^{(0,1,0)}_{-\vec\mu}
$$
\noindent
$(0,1,0)$ denotes 6-plet of $A_3$.
It is a free massive theory of six Majorana fermions:
\begin {equation}
S = \sum_{i=1}^6 \int d^2z (\psi_i \bar \partial \psi_i +
\bar \psi_i \partial \bar \psi_i + m\bar\psi_i \psi_i)
\label{eq:101}
\end {equation}
$A_4$ algebra is generated by 15 spinless charges
$$
Q^{ij}_0 = \int dz \psi_i\psi_j + \int d\bar z \bar \psi_i\bar \psi_j
$$
belonging to (1,0,1)-representation.
At spin one we have 21 conserved charges. One of them is momentum
generator
$$
p=\sum_{i=1}^6 (\int dz \psi_i\partial \psi_i -2m\int d\bar z
\bar \psi_i \psi_i)
$$
which is $A_3$-scalar. Another 20 charges form (0,2,0)-representation of
$A_3$:
$$
Q^{ij}_1(G)=\sum_{i,j=1}^6(\int dz G^{ij}\psi_i\partial\psi_j
- m\int d\bar z G^{ij} [\psi_i\bar\psi_j + \psi_j\bar\psi_i]
$$
$G$ is traceless symmetric $6\times 6$ matrix.
The charges $Q^{ij}_0$,$Q^{ij}_1$ and $\bar Q^{ij}_{-1}$
and their pairwise commutator generate $A_5^{(2)}$. Proof
is absolutely similar to two previous examples if one notes
that $A_5$ is decomposed as $(1,0,1)+(0,2,0)$ under adjoint action
of $A_3$ embedded as $so(6)$.
This algebra has been discovered independently in
\cite{sotkov}. Different twisted affine algebras appear as subalgebras
of that symmetry. Obviously, this construction is valid for any
$(21\dots 1\vert 1\dots 1)$-perturbed $WD_n$.

\section{Discussion}

We considered quantum integrable models defined as relevant perturbation
of simly-laced $W$-models ($W^{(p)}_3$, $WD^{(p)}_n$, $n\geq 3$) by operator
different from $(1,1,Adj)$ and their $p\to\infty$ limit.
All these models are generalizations of $(1,2)$-perturbed minimal CFT.
 Scattering theory of $(1,1,Adj)$-models is commonly believed
to be described in terms of finite quantum group $(X_n)_q$.
We conjecture that for fundamental weight perturbations $q$-deformation
of some twisted affine Lie algebra is the key symmetry of a theory.
This algebra is dual to affine algebra obtained as extention of
$X_n$ by a weight of perturbing operator. So we get $A_2^{(2)}$ for
$(1,2)$ perturbed CFT ($(A_2^{(2)})_q$ R-matrix  was used by F.Smirnov
in order to construct $S$-matrices for these models), $D_4^{(3)}$
for $(2,1\vert 1,1)$-perturbed $W_3$ CFT and $A_{2n-1}^{(2)}$ for
vector $(21\dots 1\vert 1\dots 1)$- perturbation of $WD_n$.
$S$-matrix construction for these algebras will be addressed in the future
paper. In the $p\to\infty$ limit $q$-symmetry becomes classical what is
explicitly demonstrated. Generators of affine twisted symmetry appear as
Noether charges of perturbed WZW models which arise in considered limit.
This result seems to be similar to recent construction of A.LeClair
\cite{lecl}. The lowest models for $n = 2,3$ can be identified as
critical Ashkin-Teller models (decoupled Isings and $Z_4$ - parafermions)
 in electric field.

\section{Acknowledgements}
I wish to thank A.Babichenko, S.Elitzur, V.Fateev, A.Rosly and
G.Sotkov for discussions. I should like to acknowledge the hospitality
of Hebrew University of Jerusalem and SISSA during the work.

This work was supported in part by Lady Davis Fellowship of Israeli
Academy of Science.


\begin{thebibliography}{99}

\bibitem{zam} A.B.Zamolodchikov,{\it JETP Lett.} {\bf 46}(1987), 160.
\bibitem{Mussardo} G.Mussardo, {\it Phys.Reports} {\bf 218}(1993), 215.
\bibitem{zamol} A.B.Zamolodchikov,{\it Adv. Studies
 Pure Math.} {\bf 19}(1989), 641.
\bibitem{Gri} G.W.Delius, M.T.Grisaru and D.Zanon,
{\it Nucl. Phys.} {\bf B382}(1992), 365.
\bibitem{corrigan}
 E.Corrigan, P.E.Dorey and R.Sasaki, preprint DTP - 93/19, YITP/U - 93-09.
\bibitem{watts}  H.G.Kausch and G.M.T.Watts,
{\it Nucl.Phys.} {\bf B386}(1992), 166.
\bibitem{EY} T.Eguchi and S.-K.Yang,
{\it Phys. Lett.} {\bf B224}(1989), 373.
\bibitem{HM} T.Hollowood and P.Mansfield, {\it Phys.Lett.} {\bf B226}
(1989), 73.
\bibitem{FF1}
B.Feigin and E.Frenkel,{\it Phys. Lett.} {\bf B276}(1992), 79.
\bibitem{FF2}
B.Feigin and E.Frenkel, preprint YITP/K - 1036.
\bibitem{FF3}
B.Feigin and E.Frenkel, preprint, November 1993.
\bibitem{IzKor} A.G.Izergin and V.E.Korepin, {\it Commun. Math. Phys.}
{\bf 79}(1981), 303.
\bibitem{Smirnov}
F.A.Smirnov,{\it Int. J. Mod. Phys.} {\bf A6}(1991), 1407.
\bibitem{ZFW}V.A.Fateev and A.B.Zamolodchikov,
{\it Nucl. Phys.} {\bf B280[FS18]}(1987), 644.
\bibitem{FLW}V.A.Fateev and S.Lukyanov,{\it JETP} {\bf 94}(1988), 23.
\bibitem{FdV}
H.J. de Vega and V.A.Fateev,{\it Int. J. Mod. Phys.} {\bf A6}(1991), 3221.
\bibitem{FatLuk} S.Lukyanov and V.A.Fateev,
preprints ITF - 88-74-76P, Kiev 1988.
\bibitem{FatPar}
V.A.Fateev,{\it Int. J. Mod. Phys.} {\bf A6}(1991), 2109.
\bibitem{Leclair}A.LeClair,{\it Phys. Lett.} {\bf B230}(1989), 103.
\bibitem{BernLec}D.Bernard and A.LeClair,
{\it Phys. Lett.} {\bf B247}(1990), 309.
\bibitem{Smir}F.Smirnov,{\it Nucl. Phys.} {\bf B337}(1990), 156.
\bibitem{SmirResh}N.Reshetikhin and F.Smirnov,
{\it Commun. Math. Phys.} {\bf 131}(1990), 157.
\bibitem{EguYang} T.Eguchi and S.-K.Yang,
{\it Phys. Lett.} {\bf B235}(1990), 282.
\bibitem{Hollow} T.J.Hollowood,
{\it Int. J. Mod. Phys.} {\bf A8}(1993), 947.
\bibitem{sotkov}G.Sotkov and M.Stanishkov,
preprint IFT - P.001/93;
\noindent
E.Abdalla, M.C.B. Abdalla, G.Sotkov and M.Stanishkov,
IFUSP-preprint-1027.
\bibitem{lecl}A.LeClair, preprint CLNS 93/1220.

\end{thebibliography}
\end{document}